\documentclass[
 reprint,
superscriptaddress,
 amsmath,amssymb,
 aps,
]{revtex4-2}

\usepackage{graphicx}
\usepackage{dcolumn}
\usepackage{bm}
\usepackage{braket}
\usepackage{orcidlink}

\usepackage{enumitem} 
\newcommand{\be}{\begin{equation}}
\newcommand{\ee}{\end{equation}}
\newcommand{\bea}{\begin{eqnarray}}
\newcommand{\eea}{\end{eqnarray}}

\newcommand{\unit}[1]{\ensuremath{\, \mathrm{#1}}}

\usepackage{hyperref} 
\hypersetup{
    colorlinks,
    linkcolor={blue!80!blue},
    citecolor={blue!80!blue},
    urlcolor={blue!80!blue}
}

\usepackage{physics}
\usepackage{float} 

\usepackage{listings}
\usepackage[usenames,dvipsnames,x11names]{xcolor}
\usepackage{tikz}

\begin{document}

\preprint{APS/123-QED}

\title{Simulating a quantum sensor: quantum state tomography of NV--spin systems}

\author{Alberto López--García\,\orcidlink{0009-0001-0023-9850}}%
 \email{alberto.lopezg@upct.es}
\affiliation{\'Area de F\'isica Aplicada, Universidad Politécnica de Cartagena member of European University of Technology EUT+, Cartagena E-30202, Spain}

\author{Aikaterini Vasilakou}%
\affiliation{\'Area de F\'isica Aplicada, Universidad Politécnica de Cartagena member of European University of Technology EUT+, Cartagena E-30202, Spain}

\author{Javier Cerrillo\ \orcidlink{0000-0001-8372-9953}}%
 \email{javier.cerrillo@upct.es}
\affiliation{\'Area de F\'isica Aplicada, Universidad Politécnica de Cartagena member of European University of Technology EUT+, Cartagena E-30202, Spain}


\begin{abstract}
We employ a quantum computer to simulate the effect of spin impurities on nitrogen-vacancy (NV) centers in diamond. As these defects operate as nanoscale quantum sensors, modeling quantum noise is crucial to identify limitations in precision. The analysis is performed by means of quantum state tomography on two transmon qubits, representing respectively the NV center and a single spin impurity, modeling either a nuclear spin or an additional NV center. We demonstrate a versatile platform to simulate benchmark protocols such as Ramsey or Hahn--echo. Although we focus on a two-spin system, the same approach opens the door to using quantum processors as scalable simulators of many--spin environments, intractable in classical simulation due to the rapid exponential growth of the Hilbert space. The results reveal the effect different spin-sensor coupling regimes have on coherence, helping to identify detection schemes that maximize the sensitivity under the effect of impurities. Moreover, the role of entanglement generation is analyzed using the Peres-Horodecki criterion and CHSH inequalities. Although no violation of the latter is observed, the presence of entanglement is confirmed. 
\end{abstract}

\maketitle


\section{\label{sec:introduction}Introduction}

Quantum simulation has been a turning point in computational physics, offering the ability to investigate and study complex dynamics in a controllable environment \cite{Cirac2012, Blatt2012, Bloch2012, Buluta2009}. As the size of the system increases, the dimension of the Hilbert space grows exponentially, making a reliable simulation on classical platforms considerably challenging \cite{Troyer2005}. In this context the idea originally proposed by Richard Feynman \cite{Feynman1982} has evolved into more than a computational strategy: it has become an experimental paradigm in which a well‑controlled quantum system is harnessed to emulate the behavior and dynamics of another quantum process.

Within the field of quantum simulation, two main approaches are typically distinguished: digital quantum simulation (DQS) and analog quantum simulation (AQS). DQS relies on decomposing the time evolution of the target system into a discrete sequence of quantum gates \cite{Lloyd1996}. However, its practical implementation remains limited by the hardware constraints of noisy intermediate--scale quantum (NISQ) devices, particularly in the absence of full error-correction protocols and given the restricted fidelity of gate implementation \cite{Preskill2018}. In contrast, AQS directly encodes the Hamiltonian of interest and continuously emulates its dynamics \cite{Georgescu2014}, often exhibiting greater robustness due to its lower overhead and reduced sensitivity to gate‑level imperfections.

Among the various platforms for realizing quantum processors, transmon-based superconducting circuits have established themselves as one of the most versatile and widely used architectures \cite{Koch2007, Blais2004}. These ``artificial atoms" offer very precise control of Hamiltonian parameters and qubit-qubit interactions in the microwave regime--the operational domain of superconducting quantum devices. This level of control has enabled the simulation of spin dynamics \cite{Salathe2015, LasHeras2014}, the exploration of circuit quantum electrodynamics of circuits (cQED) phenomena and even relativistic effects \cite{Wilson2011, Johansson2009}, as well as the implementation of quantum-sensing experiments \cite{Bylander2011, Wilen2021}-- the latter forming the focus of this work.

Quantum sensing exploits intrinsically quantum features, such as superposition and entanglement, to perform exceptionally precise measurements, enabling the detection of minute variations in physical quantities such as electric and magnetic fields \cite{Degen2017}. Among the different platforms available, the nitrogen-vacancy (NV) center in diamond stands out as a powerful platform for nanoscale sensing, particularly in nano-NMR applications \cite{Maze2008, Balasubramanian2008, Schirhagl2014, Mamin2013, Staudacher2013, LopezGarcia2025, Vetter2022, Cerrillo2021}. Despite its remarkable sensitivity, the NV center is significantly affected by magnetic‑noise sources originating from surrounding lattice spins, making the characterization of these noise mechanisms essential for accurate sensing. 

In this work, we analyze the influence of spin impurities in the vicinity of the NV center, such as carbon-13 nuclear spins, P1 centers, the intrinsic nitrogen nuclear spin of the NV defect ($N^{14}$, $N^{15}$), or electronic spins associated with unpaired electrons or neighboring NV centers \cite{Childress2006, Zhao2012, Witzel2006, Maze2008b, Hall2014, Bar-Gill2012}. To study these effects, we simulate the system using transmon-based quantum processors \cite{GarciaPerez2020, Salathe2015}, which enable the modeling of spin dynamics and impurity interactions with high controllability. Although we restrict the scope of this work to a two--qubit system, the same framework paves the way for the establishment of quantum processors as emulators of many spins, overcoming the limitations and computational capacity of classical simulation. In particular, we implement pulse sequences such as Ramsey \cite{Ramsey1950} and Hahn-echo \cite{Hahn1950}, using Gaussian-shaped MW pulses. Furthermore, we evaluate the generation of entanglement between the sensor and nearby spin by reconstructing the system's quantum state through quantum state tomography (QST) \cite{James2001, NielsenChuang}. The presence of entanglement is assessed using criteria such as the study of purity evolution, the Peres-Horodecki criterion \cite{Peres1996, Horodecki1996} and CHSH inequalities \cite{Clauser1969}. This allows us to identify non‑classical correlations across different coupling regimes and to evaluate the relevance of quantum contributions to the noise environment.

To address the goals of this work, we begin in Sect.\ref{sec:sec2} by introducing NV centers, the relevant spin impurities in their environment, and the pulse sequences employed throughout the study. Sect.\ref{sec:sec3} then details the simulation methods, including the implementation of quantum gates through shaped pulses, calibration procedures, and quantum state tomography (QST). In Sect.\ref{sec:sec4} we describe the two experimental configurations considered and discuss possible decoherence processes associated with each one. Finally, Sect.\ref{sec:sec6} presents the findings across different regimes, and Sect.\ref{sec:sec7} summarizes our conclusions.

\section{Quantum Sensing with NV Centers}
\label{sec:sec2}
In this section, we present the Hamiltonian of the NV$^-$ center and justify the effective qubit model used in this work. We then discuss how nearby spins limit sensor coherence, distinguishing regimes that can be treated as classical perturbations from those that involve coherent quantum interactions. Finally, we summarize the control protocols employed in this work, focusing on Ramsey and Hahn--echo sequences.

\subsection{\label{sec:level3}NV Centers and Spin Environments}


The NV center has become one of the most promising quantum sensors because of its ability to detect minute electric and magnetic fields on the nanoscale, even at room temperature \cite{Rondin2014, Wolf2015, Taylor2008, Acosta2009, Barry2020}. Structurally, an NV center is a defect in the diamond's crystal structure, where two carbon atoms are substituted by a single nitrogen atom, producing a system with different electronic spin states depending on its charge. The negatively charged NV center (NV$^-$) features a ground spin triplet state ($S=1$), whose sub-levels ($m_s = 0, \pm 1$) are sensitive to external fields via Zeeman splitting \cite{Doherty2013, Gruber1997, Jelezko2004, Balasubramanian2009, Ishikawa2012}. Moreover, the spin can be initialized in the state $\ket{m_s=0}$ and read out via spin-dependent fluorescence using ODMR techniques. The associated Hamiltonian is
 \begin{equation} 
    H = DS_z^2 + \mu B S_z, 
\end{equation}
where $D/2 \pi \approx 2.87 \unit{GHz}$ is the zero-field splitting between state $\ket{0}$ and states $\ket{\pm1}$, $\mu$ is the gyromagnetic ratio, and $S_z$ the spin operator along the NV axis.
In this work, we adopt an effective two--level description of the NV center by selecting a single transition within the $S=1$ ground--state manifold under an external magnetic field that lifts the $\ket{m_s = \pm 1}$ degeneracy. This selection is consistent with the qubit--level scheme of transmon--based processors and is sufficient to study coherence protocols and different sensor--impurity interactions. However, we note that control and entanglement protocols using the three--level manifold of the NV center are an increasingly active area of research \cite{Vetter2022, Cerrillo2021, LopezGarcia2025, MorillasRozas2025, Neumann2010, Taminiau2014, Waldherr2014, Bradley2019}.

\begin{figure}[h!]
\centering
\includegraphics[width=\columnwidth]{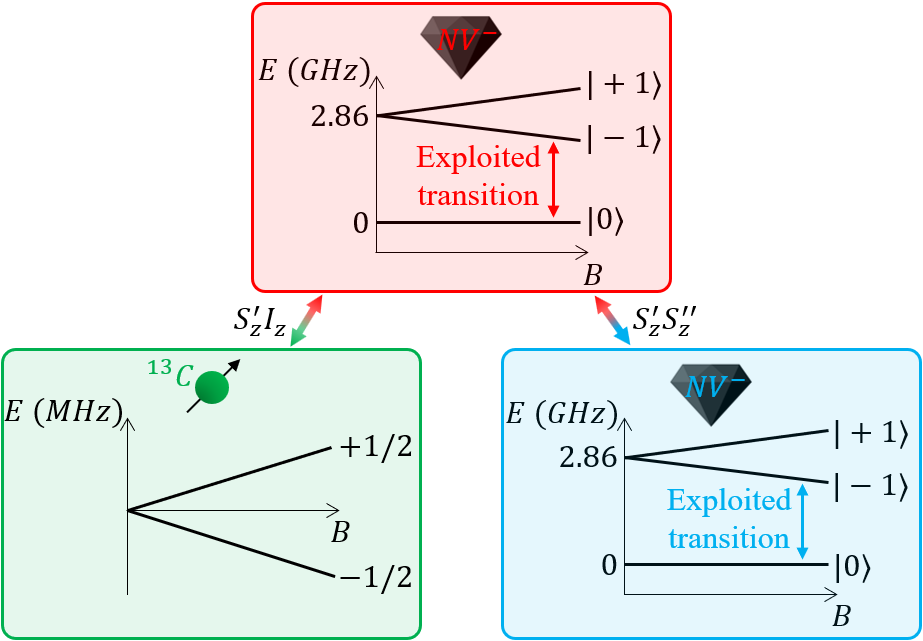}
\caption{Schematic of sensor--impurity configurations considered in the present work. (Top, red) Main NV$^{-}$ considered as the sensor, where the $\ket{0} \leftrightarrow \ket{-1}$ is exploited for sensing. (Red--green combination) NV$^{-}$--$^{13}$C coupling described by the interaction term $S'_{z}I_{z}$, representing a nuclear spin $I = 1/2$ that induces stochastic magnetic--field noise in the NV$^{-}$'s dynamics. (Red--blue combination) NV$^{-}$--NV$^{-}$ coupling described by the interaction term $S'_{z}S''_{z}$, corresponding to a coherent regime that induces quantum correlations between both NV centers.}
\label{fig:ESQUEMA}
\end{figure}

In NV-based quantum sensing, nearby spins limit coherence by imprinting fluctuating phases on the sensor dynamics. Depending on the coupling strength and relevant timescales, this effect can be described either as classical dephasing --uncorrelated stochastic fluctuations with the sensor state-- or as coherent quantum dynamics emerging from the joint evolution of sensor and impurity \cite{Reinhard2012}. In the latter regime, nonclassical correlations and even entanglement between both subsystems may arise. In an attempt to explore both possibilities, we will consider two forms of impurities (see Fig. \ref{fig:ESQUEMA}):

\paragraph{Nuclear spins.}

Nuclear spins surrounding the NV center such as carbon-13 nuclear spins, P1 centers or the intrinsic nitrogen nuclear spin of the NV defect ($N^{14}$, $N^{15}$) typically act as weakly coupled magnetic impurities. Since their dynamics are slow compared to the NV electronic spin, their main effect is to shift the local field and induce quasi--static dephasing, rather than coherent exchanges. In this regime, one expects a reduction in coherence with no significant sensor--impurity entanglement.

\paragraph{Nearby NV as an impurity.}

In contrast, a nearby NV center will react to the same MW pulses as the target NV and is in the same dynamical regime. Therefore, its dipolar coupling can produce coherent sensor--impurity evolution and generate quantum correlations, including entanglement.

\subsection{Ramsey and Hahn echo sequences}

\begin{figure}[h!]
\centering
\includegraphics[width=0.9\columnwidth]{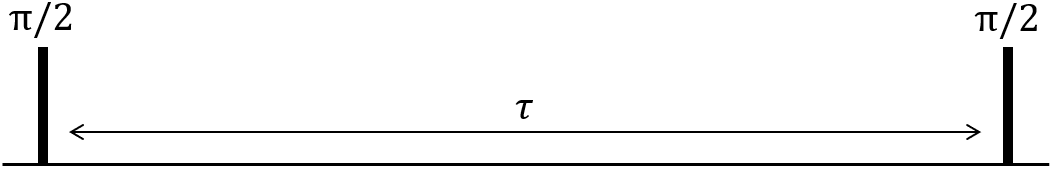}
\caption{Ramsey sequence used to estimate $T_2^*$. Two $\pi /2$ pulses are separated by a free--evolution time $\tau$.}
\label{fig:secuenciaRamsey}
\end{figure}

In open quantum systems such as the NV center, the dynamics is generally non--unitary due to interactions with the environment. This leads to decoherence, commonly quantified by the dephasing time $T_2^*$, which is measured from a Ramsey experiment (see Fig. \ref{fig:secuenciaRamsey}) \cite{Ramsey1950} implemented as follows:

\begin{enumerate}
    \item Initialize the qubit in the ground state $\ket{0}$.
    \item Apply a $\pi/2$ pulse to prepare a coherent superposition.
    \item Let the system evolve for a time $\tau$, during which a relative phase is accumulated due to environmental interactions.
    \item Apply a final $\pi/2$ pulse to map the accumulated phase into the measurement basis.
    \end{enumerate}

\begin{figure}[h!]
\centering
\includegraphics[width=0.9\columnwidth]{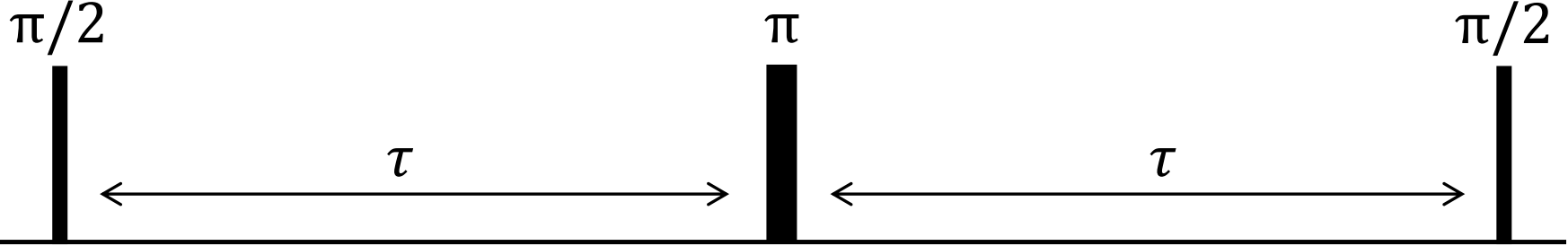}
\caption{Hahn echo sequence used to estimate $T_{2}$. A refocusing $\pi$ pulse is implemented between the two free--evolution intervals.}
\label{fig:secuencia Hahn echo}
\end{figure}

This decoherence can be partially compensated by using a Hahn--echo sequence (see Fig. \ref{fig:secuencia Hahn echo}) \cite{Hahn1950, Carr1954, Meiboom1958}, implemented as follows:
    
\begin{enumerate}
    \item Initialize the qubit in the ground state $\ket{0}$.
    \item Apply a $\pi/2$ pulse to prepare a coherent superposition.
    \item Let the system evolve for a time $\tau$, during which phase is accumulated due to environmental interactions.
    \item Apply a $\pi$ pulse to invert the spin.
    \item Let the system evolve for a second time $\tau$, refocusing the phase accumulated during the first interval and producing an \textit{echo}.
    \item Apply a final $\pi/2$ pulse to map the accumulated phase into the measurement basis.
    \end{enumerate}

The central $\pi$ pulse refocuses the phase evolution, effectively extending the coherence time of the qubit. As a result, Hahn--echo yields a longer characteristic time, denoted by $T_{2}$ \cite{deLange2010, Stanwix2010, Uhrig2007, Souza2012, Alvarez2011}. 

\section{Experimental Method}
\label{sec:sec3}

\begin{figure}[h!]
\centering
\includegraphics[width=0.4\textwidth]{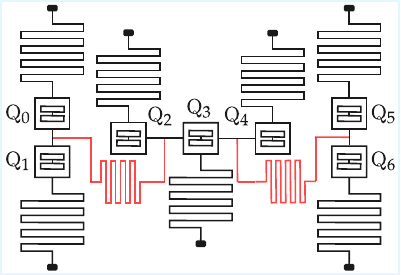}
\caption{Seven-qubit superconducting architecture based on transmon qubits. Each qubit is coupled to an individual resonator for control and readout, while two bus resonators (red) enable coherent two-qubit gates.}
\label{fig:arquitectura}
\end{figure}

To simulate the interaction between the NV center and nearby spin impurities, we implement the protocol on a transmon--based superconducting quantum processors accessed via IBM Quantum Lab \cite{Qiskit2019, Alexander2020}. In particular, we use the devices \textit{Manila} and \textit{Nairobi}, featuring 5 and 7 qubits (see Fig. \ref{fig:arquitectura}), respectively \cite{GarciaPerez2020}. The NV-impurity system is modeled as a pair of coupled qubits, enabling us to explore different coupling regimes and control sequences such as Ramsey and Hahn--echo. Throughout this work, qubit 0 (Q$_{0}$) represents the NV sensor, while the qubit 1 (Q$_{1}$) represents the impurity. Although we focus on a two--qubit system (sensor--impurity), the same pulse--level strategy can be extended to larger systems to emulate multi-spin environments. It allows controlled studies of spin baths, which are limited from a classical computing perspective. 

The following subsections describe the Gaussian pulse--based gate implementation, the calibration procedure and the quantum state tomography (QST) protocol used to extract the relevant information.

\subsection{\label{sec:level2}Gaussian Pulse-Based Quantum Gates}

\begin{figure}[h!]
\centering
\includegraphics[width=\columnwidth]{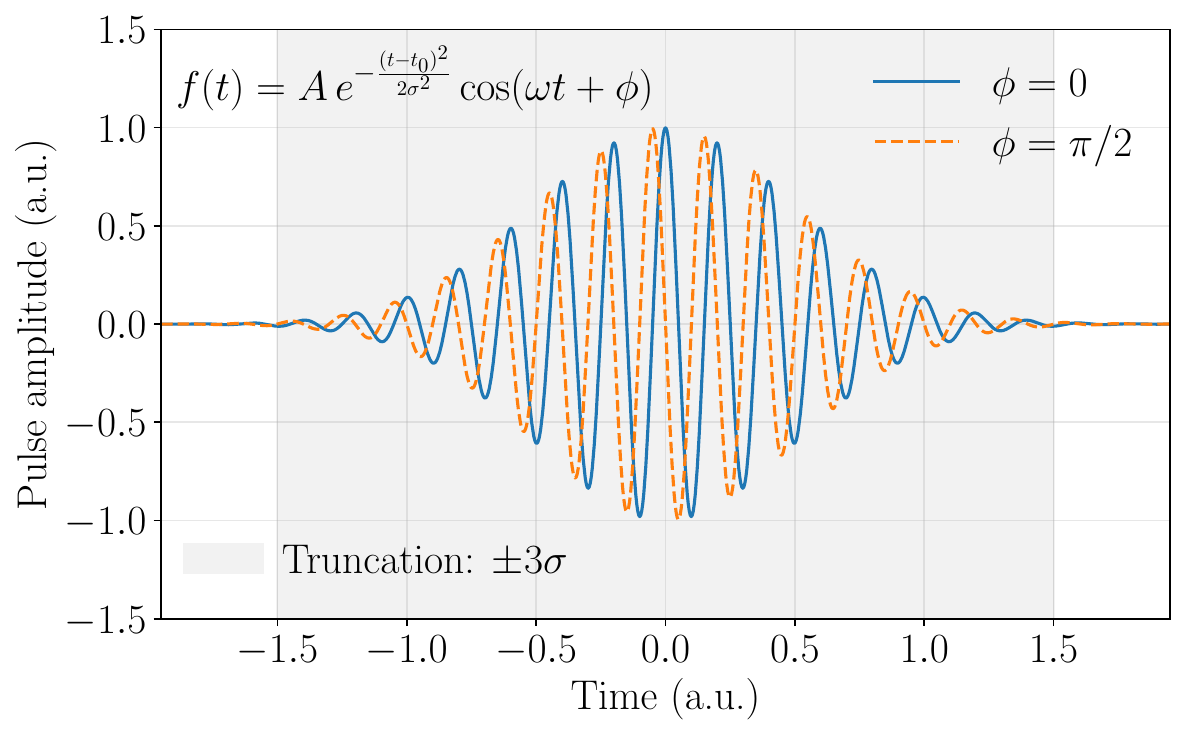}
\caption{Gaussian-shaped microwave pulses for coherent qubit manipulation, illustrating two rotation axes determined by the phase $\phi$. The orange solid curve corresponds to a pulse with $\phi=0$ (rotation around $X$) and the dashed blue curve corresponds to $\phi=\pi/2$ (rotation around $Y$). The shaded gray region denotes the truncated window where the pulse is applied ($\pm 3\sigma$); outside this window the pulse is assumed negligible. The pulses are characterized by amplitude ($A$), angular frequency ($\omega$), phase ($\phi$), temporal width ($\sigma$), and pulse center ($t_0$).}
\label{fig:gaussian_pulses}
\end{figure}

Precise quantum state manipulation is essential in quantum sensing and is implemented in superconducting qubit platforms via shaped microwave pulses \cite{Krantz2019, Blais2021}. Mathematically, a Gaussian-shaped microwave (MW) pulse is represented as
\begin{equation}
    f(t) = A\,e^{-\frac{(t - t_0)^2}{2\sigma^2}} \cos(\omega t + \phi)
\end{equation}
where $A$ denotes the amplitude, $t_0$ is the pulse center, $\sigma$ represents the pulse width controlling the temporal range, $\omega$ is the angular frequency resonant with the targeted transition, and $\phi$ defines the pulse phase. In practice, the pulse is truncated to a finite duration for experimental implementations (see Fig. \ref{fig:gaussian_pulses}).

Gaussian pulses enable precise rotations on the Bloch sphere, a conceptual model to represent quantum state dynamics of a two level system \cite{Bloch1946}. By selecting specific parameters, particularly amplitude and duration, we perform rotations by angles such as $\pi$ and $\pi/2$, essential for quantum gate operations like spin initialization, coherent manipulation in NV-center-based sensors and readout \cite{NielsenChuang, Motzoi2009}. For instance, a pulse that yields a rotation of angle $\theta$ around an axis defined by the phase $\phi$ can be expressed in terms of the unitary operator
\be
U(\theta, \phi) = e^{-i \frac{\theta}{2} \left( \sigma_x \cos\phi + \sigma_y \sin\phi \right)}
\ee
with $\sigma_x$ and $\sigma_y$ being the Pauli matrices. Therefore, a pulse with $\phi= 0$ and angle $\pi/2$ implements the operator $U(\pi/2, 0) = e^{-i\frac{\pi}{4}\sigma_x}$, producing rotations around the $X-$axis, crucial for Hahn-echo sequences and dynamical decoupling protocols for quantum sensing \cite{Ryan2010, deLange2010, Wang2019}. 


\subsection{\label{sec:level4}Calibration}

Qubit calibration is a crucial part of the experimental procedure, made up by several steps that fully characterize the device.

\subsubsection{Frequency sweep}

\begin{figure}[h!]
\centering
\includegraphics[width=0.48\textwidth]{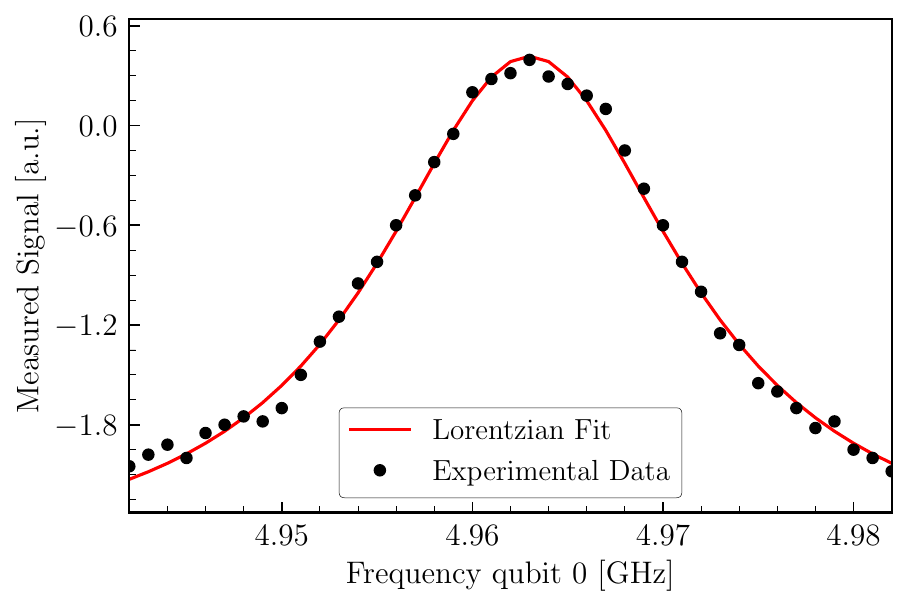}
\caption{Experimental frequency sweep performed on qubit 0. The measured data (black dots) show a clear resonance peak. A Lorentzian fit (red line) is used to determine the qubit resonance frequency, yielding $f \approx 4.962\,\mathrm{GHz}$.}
\label{fig:imagen10}
\end{figure}

The resonance frequency of a qubit is set by the energy splitting between its quantum states $\ket{0}$ and $\ket{1}$. Experimentally, we first estimate the qubit frequency using a frequency sweep around the nominal operating value of $5\,\mathrm{GHz}$ \cite{Koch2007}. This consists of varying the microwave drive frequency and measuring the qubit response (e.g., excited--state population or signal intensity). The response exhibits a resonance peak, from which the qubit frequency can be extracted via fitting (see Fig. \ref{fig:imagen10}). 

Although frequency sweeping provides a useful initial reference, the precision may be improved via Ramsey spectroscopy. For this, the Rabi frequency of the MW pulses must be determined.

\subsubsection{Rabi oscillations}

Once the resonance frequency has been estimated, the next step is to identify the drive amplitude required to induce coherent transitions between $\ket{0}$ and $\ket{1}$. This allows us to implement a $\pi$ pulse (i.e., an $X$ gate) that inverts the population.

\begin{figure}[h!]
\centering
\includegraphics[width=0.49\textwidth]{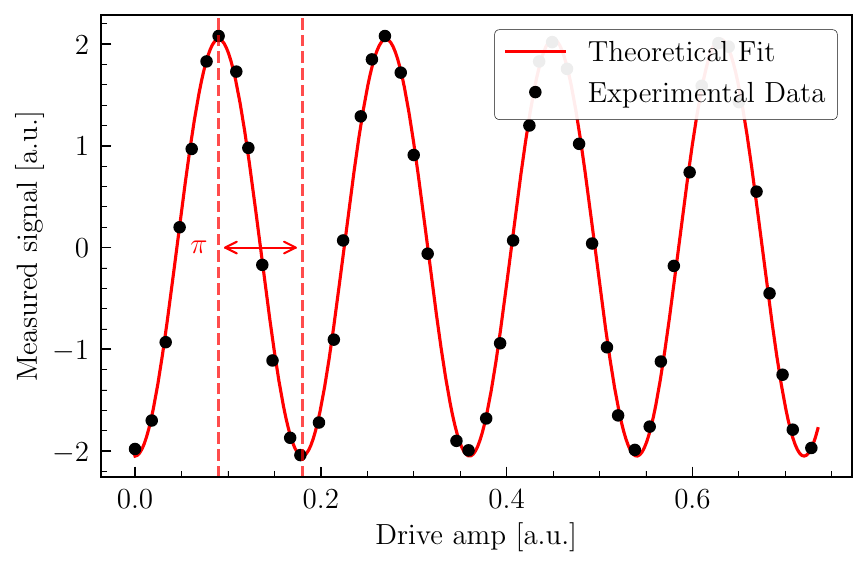}
\caption{Amplitude sweep showing Rabi oscillations for qubit 0. The measured data (black dots) are fitted with a cosine function (red line) to extract the $\pi$--pulse amplitude, yielding $A \approx 0.095 \mathrm{a.u.}$.}
\label{Amplitud de Rabi}
\end{figure}

In this procedure, the drive frequency is fixed at the estimated resonance frequency while the drive amplitude is varied. For each amplitude, we measure the qubit response and observe oscillations in the signal intensity, known as ``Rabi oscillations'' \cite{Rabi1937}, which reflect periodic transitions between $\ket{0}$ and $\ket{1}$ under continuous driving (see Fig. \ref{Amplitud de Rabi}). The experimental data follow a sinusoidal dependence and are fitted with a cosine function to characterize the oscillations. The relevant value is the amplitude corresponding to a $\pi$ rotation. Once the driving amplitude A for a $\pi$ pulse is determined, other rotations can be implemented by rescaling this amplitude; for instance, a $\pi/2$ pulse is applied using $A/2$.

\subsubsection{Readout Discriminator}

The next step is to calibrate the readout discriminator, which is crucial for reliable assignment of the qubit state. This procedure relies on measuring the complex response in the IQ--plane, where the in-phase (I) and quadrature (Q) components correspond to the real and imaginary parts of the readout signal, respectively \cite{Gambetta2007, Walter2017}. 

\begin{figure}[h!]
\centering
\includegraphics[width=0.49\textwidth]{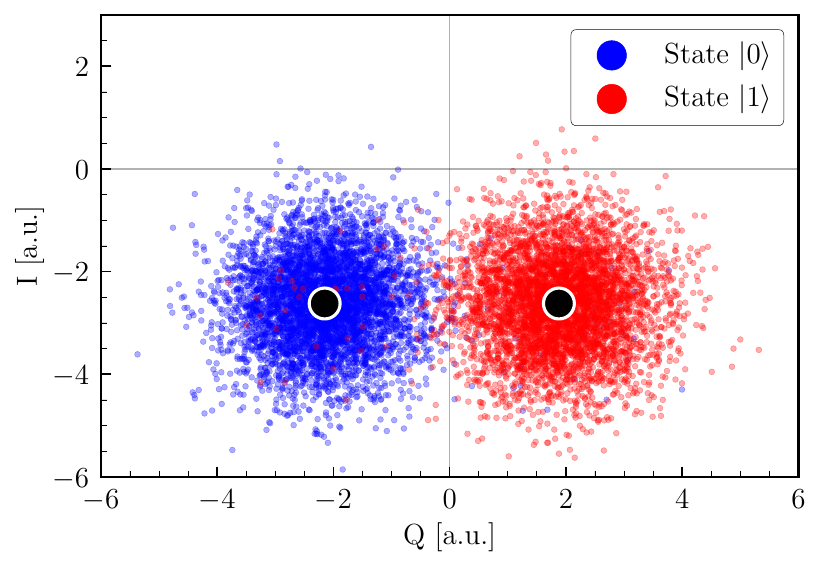}
\caption{Experimental discriminator between $\ket{0}$ and $\ket{1}$ in the IQ-plane. Blue and red points correspond to repeated measurements of a qubit prepared in $\ket{0}$ and $\ket{1}$, respectively. Black dots indicate the reference points used for discrimination.}
\label{fig:imagen12}
\end{figure}

We prepare the well-defined states $\ket{0}$ and $\ket{1}$ and perform repeated readout. Each measurement yields a point in the IQ plane whose location depends on the qubit state. The resulting data form two distinct distributions (see Fig.~\ref{fig:imagen12}), from which a linear classifier can be defined to optimally split both distributions. This discrimination procedure allows assigning each measurement outcome to either state $\ket{0}$ or $\ket{1}$ based on its position in the IQ plane. 

\subsubsection{Precise Frequency via Ramsey Protocol}

\begin{figure}[h!]
\centering
\includegraphics[width=0.49\textwidth]{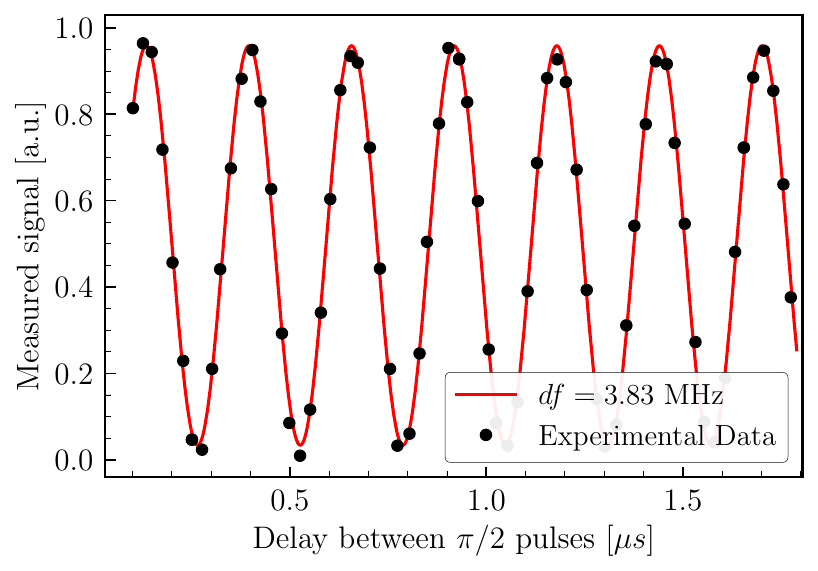}
\caption{Ramsey measurement on qubit 0 to determine the resonance frequency precisely. The measured data (black dots) are fitted with a sinusoidal function (red line) to extract the detuning frequency of the qubit, yielding $df \cong 3.83 \mathrm{MHz}$.}
\label{fig:imagen13}
\end{figure}

As a final calibration step, we determine the qubit resonant frequency more precisely using a Ramsey sequence \cite{Ramsey1950}, which is sensitive to detuning between the applied and the resonance frequency. The Ramsey sequence (Fig.\ref{fig:secuenciaRamsey}) consists of two $\pi/2$ pulses separated by a free-evolution time. During this interval, a relative phase is accumulated when the applied frequency differs from the resonance one, producing oscillations in the measured population with frequency proportional to the detuning (see Fig. \ref{fig:imagen13}). By fitting these oscillations, we estimate the detuning and adjust the pulse frequency accordingly.  

\subsection{\label{sec:level5}Quantum State Tomography}

Quantum state tomography (QST) is a standard tool to fully characterize a quantum state by reconstructing its density matrix $\rho$, which contains both populations and coherences \cite{NielsenChuang, James2001, Bruning2012}. Although the density matrix of a two-qubit-system contains 16 real parameters, we implement a procedure based on performing 9 measurements in different bases, specifically Pauli bases $X$, $Y$ and $Z$ for each of the two qubits
\begin{align*}
     \{& X\otimes X, X\otimes Y, X\otimes Z,\\ &Y\otimes X, Y\otimes Y, Y\otimes Z,\\ 
     &Z\otimes X, Z\otimes Y, Z\otimes Z\}.
\end{align*}
Each combination provides an expectation value $\langle \sigma_i \otimes \sigma_j \rangle$, which collectively enables the reconstruction of $\rho$ \cite{Smolin2012, Gross2010, BlumeKohout2010, Cramer2010}.

\begin{figure}
\centering
\includegraphics[width=\columnwidth]{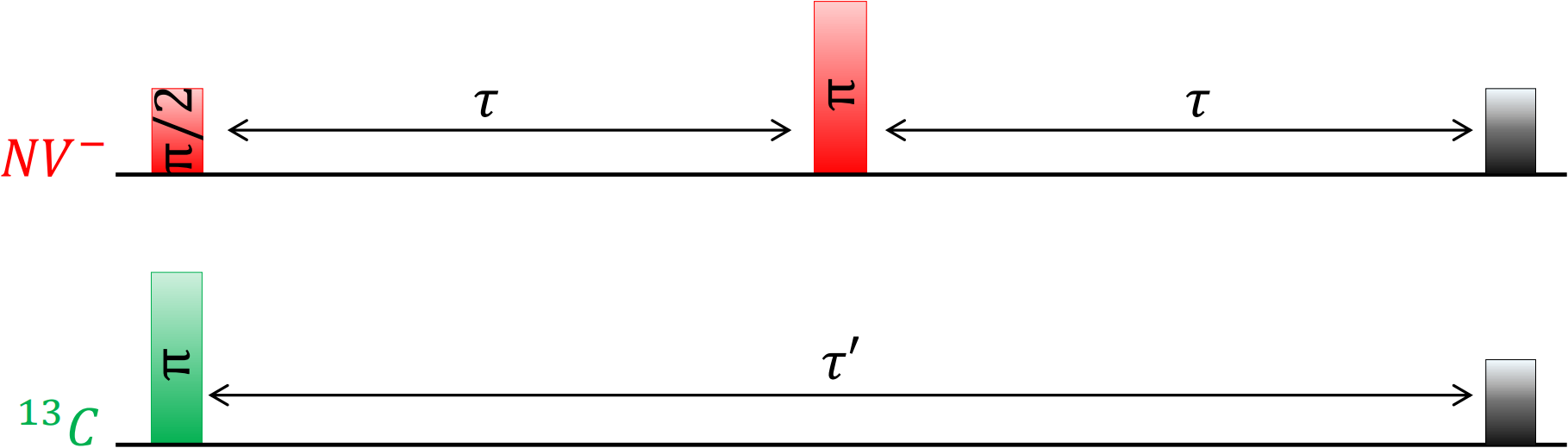}
\caption{Pulse--sequence schedule for the NV$^{-}$--$^{13}$C configuration. The sensor (red, qubit 0) undergoes a Hahn--echo. The impurity (green, qubit 1) is driven by a $\pi$ pulse to prepare an excited state and allow relaxation process during the evolution, producing dynamic dephasing on the sensor. Black blocks represent the measurement.}
\label{fig:secuencia1}
\end{figure}

\begin{figure}
\centering
\includegraphics[width=\columnwidth]{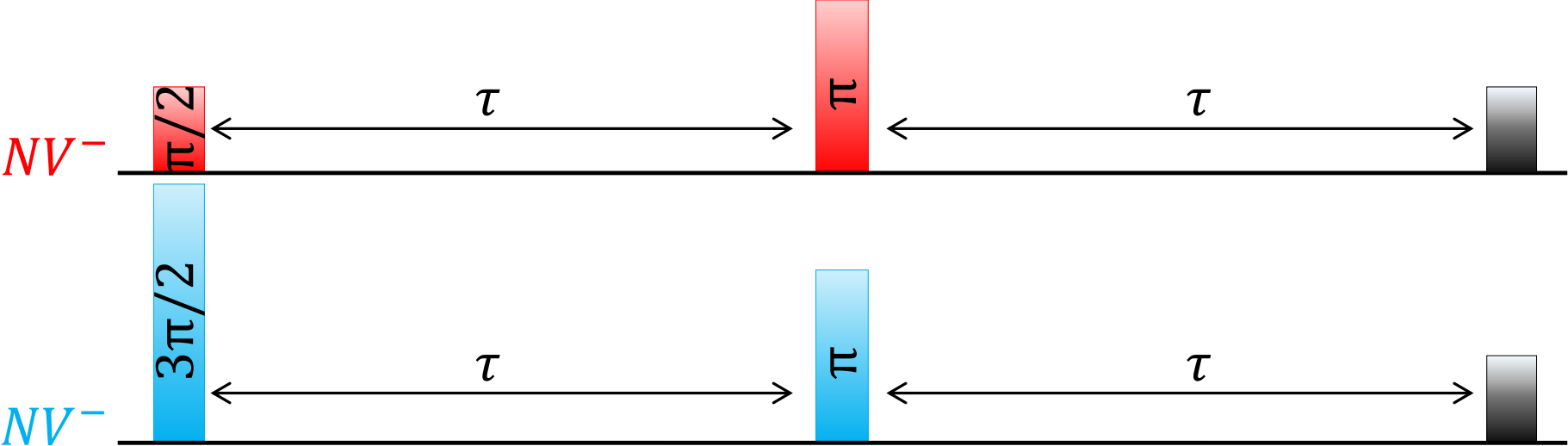}
\caption{Pulse--sequence schedule for the NV$^{-}$-- NV$^{-}$ configuration. Both qubits are driven with Hahn--echo sequences. The sensor (qubit 0) is initialized in state $\ket{+}$, while the impurity (qubit 1) is initialized in state $\ket{-}$ to induce coherent sensor--impurity dynamics. Black blocks represent the measurement.}
\label{fig:secuencia2}
\end{figure}

\subsection{Quantum Simulation Sequences}
\label{sec:sec4}

In order to simulate the effect of environmental spins in the context of a quantum sensing experiment with an NV center, we implement two different instances of a two--qubit protocol: qubit 0 represents the NV sensor where a Hahn echo sequence is being implemented and qubit 1 represents the impurity.

Under a Hahn‑echo sequence, an NV center is perturbed only by dynamical noise. The nearby nuclear spins (such as $^{13}$C) are largely unaffected by the microwave pulses that drive the NV center, so their dynamics must originate from a different mechanism—for instance, nuclear‑spin relaxation. In this regime, dipolar or hyperfine interactions between the NV center and the nucleus are expected to influence the measured $T_2$ time. In order to simulate this effect, we resort to the pulse sequence in Fig. \ref{fig:secuencia1} where the NV sensor (qubit 0) is driven with a Hahn--echo sequence, while the impurity (qubit 1) is prepared with an initial $\pi$ pulse to initiate a relaxation process.

When the impurity is another NV center in the vicinity, it is expected to be sensitive to MW pulses directed to the original NV center. We represent the situation of two parallel NV centers with the sequence in Fig. \ref{fig:secuencia2}, where both qubits are driven with a Hahn--echo protocol. However, the sensor (qubit 0) is initialized in the state $\ket{+}$, while the impurity (qubit 1) is initialized in the state $\ket{-}$, to maximize the relative phase evolution and improve the observation of coherent oscillations and correlations in the system induced by the coupling. In this setting, the dipolar interaction between both NV centers induces a coherent evolution of the form $\cos{At}\ket{+-}+i\sin{At}\ket{-+}$, where $A$ is the strength of the dipolar interaction. For times $t=k\pi/4A$ entanglement is expected between both NV centers, with its quality being reduced by decoherence effects affecting both NV centers.

These two instances are but a representation of the rich phenomenology that may be explored with transmon-based simulation of quantum sensing experiments.

\begin{figure}
\centering
\includegraphics[width=0.5\textwidth]{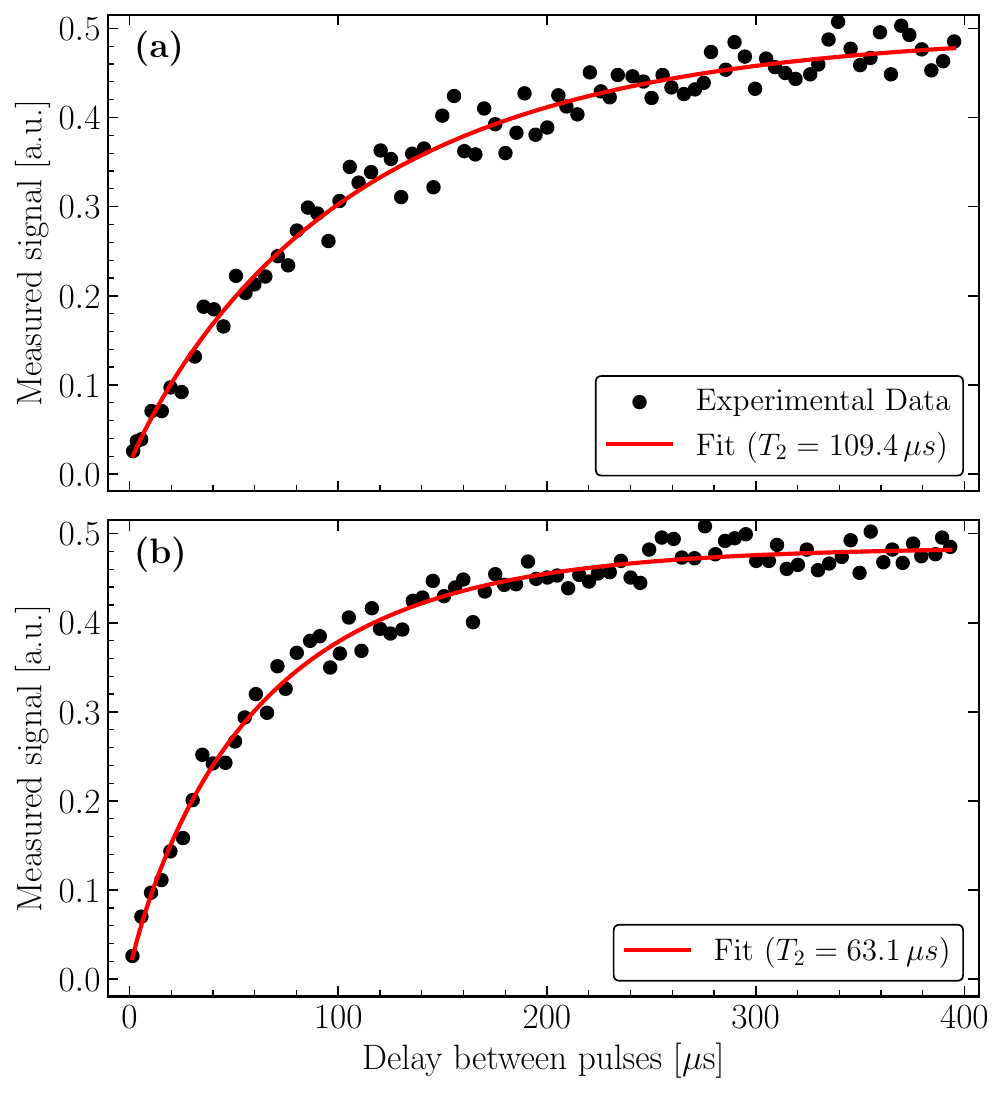}
\caption{$T_{2}$ decay of qubit 0 (sensor) under different impurity configurations. Black dots represent experimental data, while the red line is the fitting to extract the coherence time $T_{2}$. \textbf{(a)} Natural decoherence without perturbation of surrounding impurities. \textbf{(b)} Nuclear spin (qubit 1) prepared in the excited state, allowing relaxation and inducing additional decoherence of the sensor.}
\label{fig:t2_bachelor}
\end{figure}

\section{Experimental results and Analysis}
\label{sec:sec6}

\begin{figure}
\centering
\includegraphics[width=0.5\textwidth]{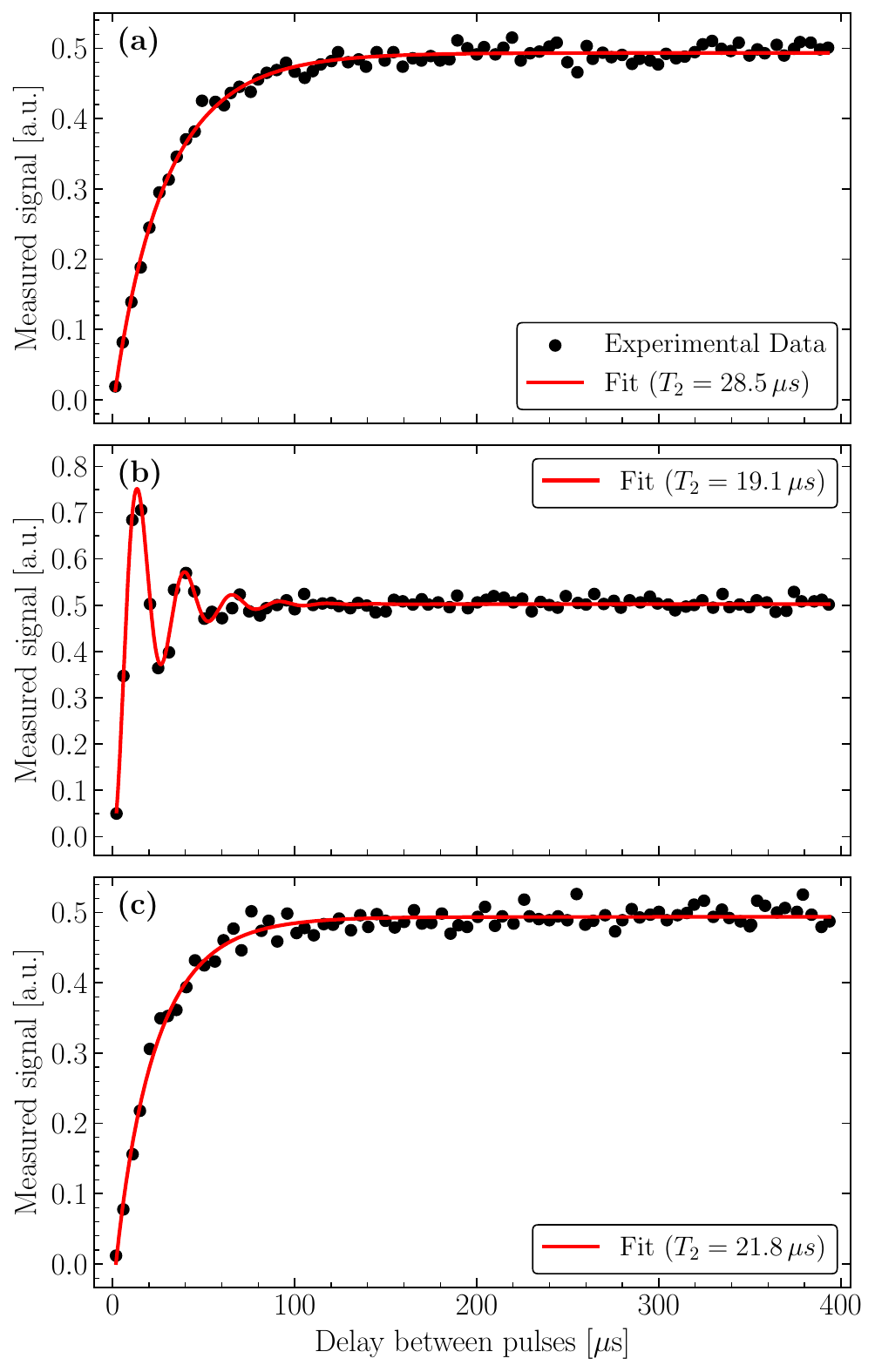}
\caption{$T_{2}$ decay of qubit 0 (sensor) under different impurity configurations. Black dots correspond to the experimental data, while the red line is the fitting to extract the coherence time $T_{2}$. \textbf{(a)} Natural decoherence without perturbation of surrounding impurities. \textbf{(b)} NV--NV configuration with a Hahn--echo sequence applied to qubit 1, where coherent oscillations are observed. \textbf{(c)} Impurity sequence analogous to a nuclear spin.}
\label{fig:t2_master_all}
\end{figure}

\subsection{Decoherence}
\label{sec:decoherence}

We first characterize the natural coherence time $T_{2}$ of the sensor qubit (qubit 0) and then repeat the measurement under each impurity configuration.

For the NV--nuclear spin configuration, the natural coherence time of qubit 0 is $T_{2} = 109.40\,\mu\text{s}$ (see Fig. \ref{fig:t2_bachelor}. \textbf{(a)}). When coupling to qubit 1, $T_{2}$ is reduced to $63.10\,\mu\text{s}$ (see Fig. \ref{fig:t2_bachelor}. \textbf{(b)}). We attribute this reduction to the relaxation process of the spectator qubit (qubit 1), which dephases the sensor and reduces its coherence stochastically in a process known as spectator-decay induced dephasing (SDID) \cite{Jurcevic2022}.

For the NV--NV configuration, the natural coherence time of qubit 0 is $T_{2} = 28.50\,\mu\text{s}$ (see Fig. \ref{fig:t2_master_all}. \textbf{(a)}). When coupling to qubit 1 driven with Hahn--echo sequence, $T_{2}$ is reduced to $19.10\,\mu\text{s}$ and observe oscillations in the coherence decay (see Fig. \ref{fig:t2_master_all}. \textbf{(b)}). For comparison, using an impurity sequence analogous to Fig. \ref{fig:secuencia1} yields $T_{2} = 21.80\,\mu\text{s}$ (see Fig. \ref{fig:t2_master_all}. \textbf{(c)}). Beyond the SDID effect, the oscillatory behavior suggests the presence of coherent dynamics and possibly entanglement \cite{Dolde2013, Bernien2013, Pfaff2014}, in contrast to the previous nuclear--spin case. To clarify the origin of these effects, we analyze the QST data to assess sensor--impurity correlations and entanglement in both configurations.

\begin{figure}
\centering
\includegraphics[width=0.5\textwidth]{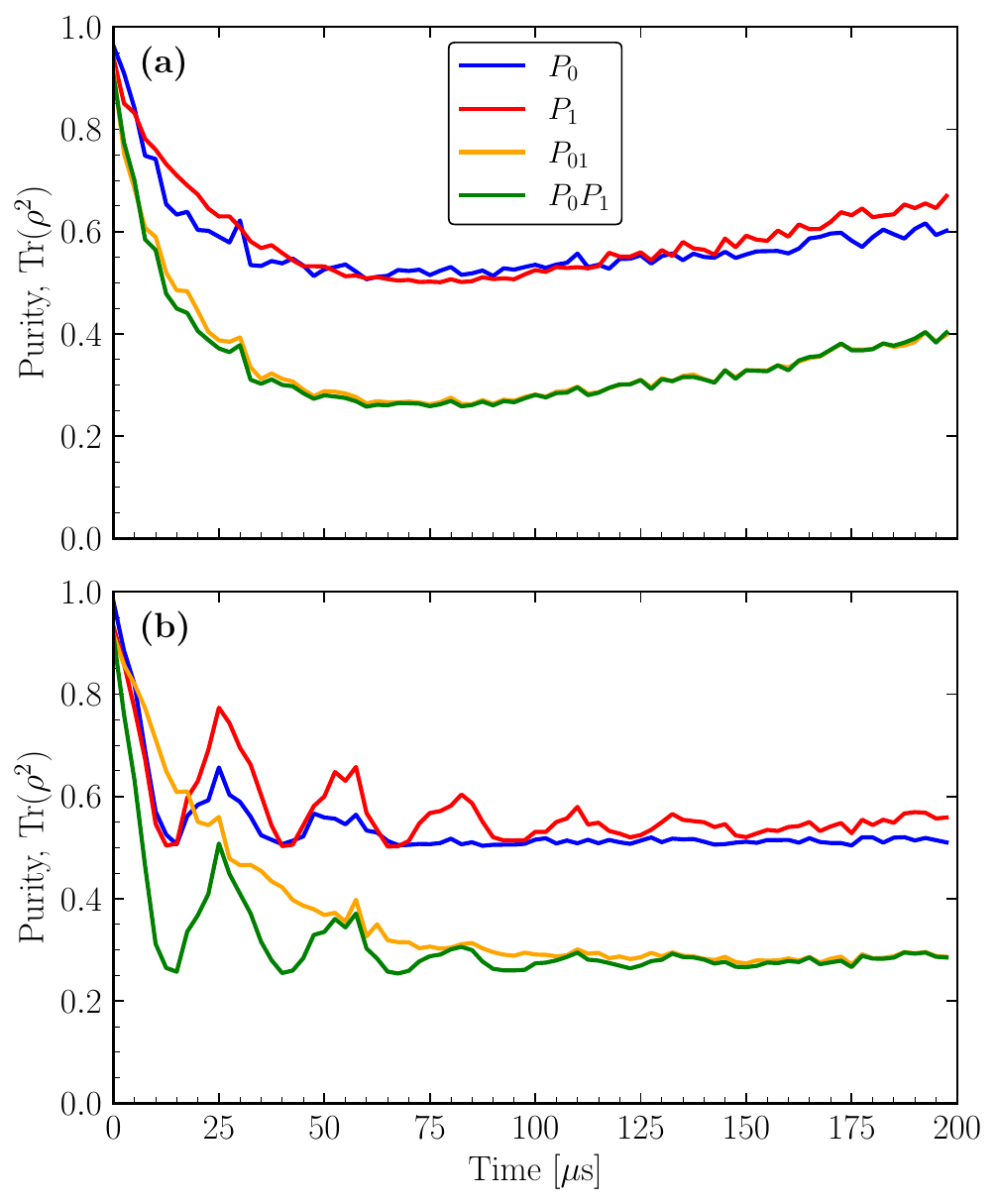}
\caption{Purity dynamics over time for sequence 1 \textbf{(a)} and sequence 2 \textbf{(b)}. Each line represents the evolution of four different purities: the purity of qubit 0 ($P_{0}$, blue), the purity of qubit 1 ($P_{1}$, red), the purity of the joint state ($P_{01}$, orange) and the product of individual purites ($P_{1}P_{0}$, green).}
\label{fig:pureza}
\end{figure}

\subsection{Purity}

We begin by reconstructing the reduced density matrices of the sensor and the impurity. These matrices encode the corresponding Bloch vectors and allow us to quantify the purity of each qubit. The purity is defined as $P=\mathrm{Tr}(\rho ^2)$ with $P=1$ for a pure state and $P<1$ indicating mixedness due to dephasing and/or decoherence \cite{NielsenChuang}. For a two--qubit state, we compute $P_{01}=\mathrm{Tr}(\rho_{01} ^2)$ and compare it with the product $P_{0}P_{1}$ of the purities of each qubit; deviations between these quantities indicate correlated sensor--impurity dynamics.

For the case of an NV center coupled to a nuclear spin, the upper panel of Fig. \ref{fig:pureza} shows an initial drop in purity due to decoherence and environmental noise. However, as the system evolves, the purity values stabilize and reach a steady state. Importantly, $P_{01}$ remains close to $P_{0}P_{1}$ throughout the evolution, indicating weak correlations and no evidence of entanglement in this regime.
In contrast, for the case of an NV center coupled to another NV center, the lower panel of Fig. \ref{fig:pureza} exhibits a clear deviation between $P_{01}$ and $P_{0}P_{1}$ over a significant part of the evolution, consistent with correlated and non--classical dynamics.

\begin{figure}
\centering
\includegraphics[width=0.5\textwidth]{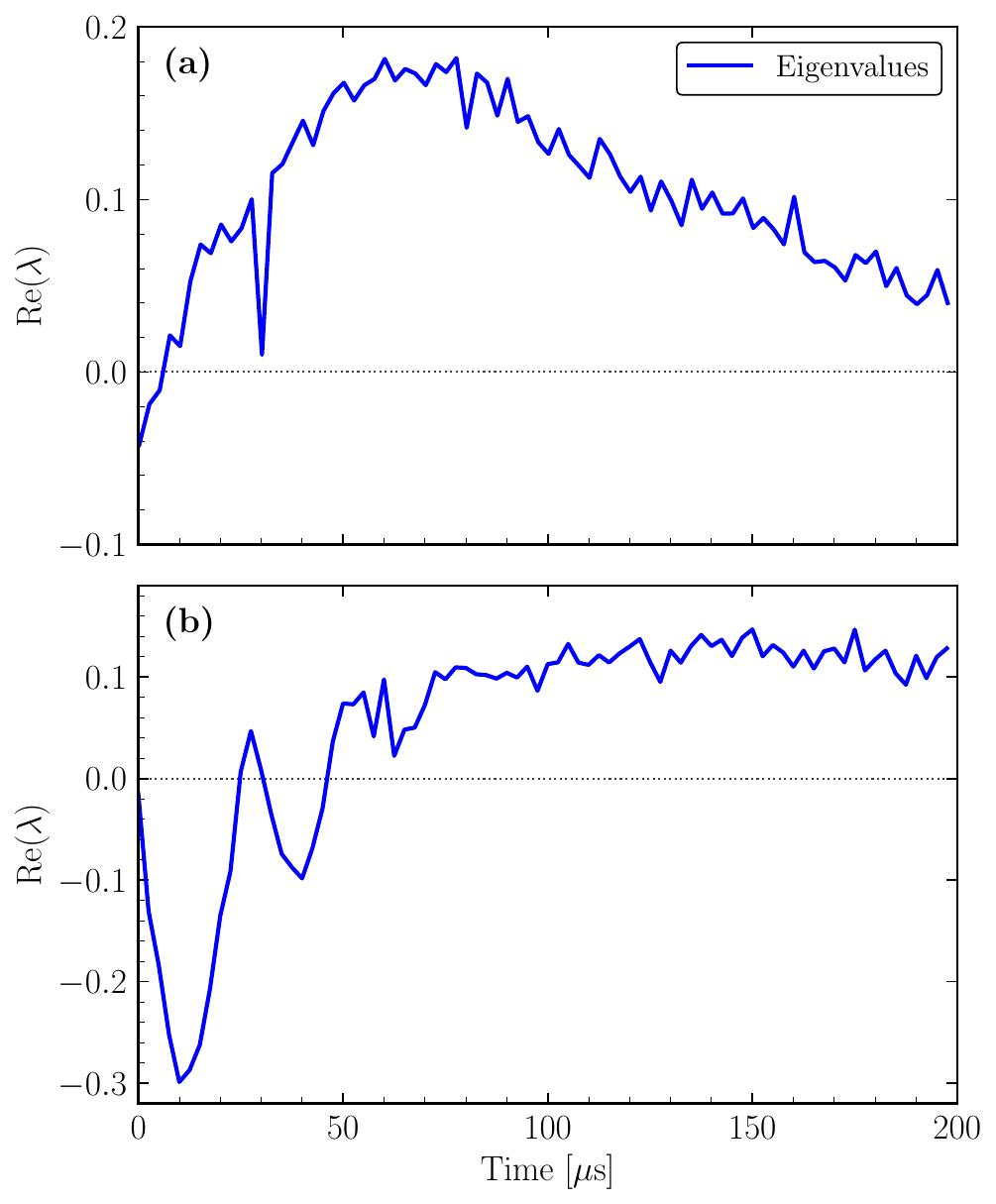}
\caption{Temporal evolution of the minimum eigenvalue of the partially transposed density operator for \textbf{(a)} an NV center coupled to an (\(I = 1/2\)) nuclear spin and \textbf{(b)} two coupled NV centers. The gray--dotted horizontal line at zero represents the threshold where non--negative eigenvalues, apart from the spurious initial ones, indicate the absence of entanglement, meaning that the state evolves towards a separable condition.}
\label{fig:autovalores}
\end{figure}

\subsection{Peres--Horodecki criterion}

We next apply the Peres--Horodecki criterion \cite{Peres1996, Horodecki1996, Horodecki2009, Vidal2002}, also known as positive partial transpose (PPT) criterion, to determine whether the two--qubit state is separable or entangled. We compute the partial transpose of $\rho_{01}$ with respect to one subsystem; the presence of at least one negative eigenvalue certifies entanglement. The result is independent from the subsystem used to compute the transpose since, in general,
$\rho^{T_A}=(\rho^{T_B})^T$. For the case of an NV center coupled to a nuclear spin (upper panel of Fig.\ref{fig:autovalores}) all eigenvalues remain non--negative except for a few initial values. This points to the absence of entanglement. In contrast, for the NV center coupled to another NV center, the lower panel of Fig.\ref{fig:autovalores} shows negative eigenvalues at the beginning of the evolution, indicating entanglement according to the PPT criterion. At longer times, all eigenvalues become positive, consistent with a transition to a separable state dominated by decoherence.

We note that the initial data points exhibit a small negative offset in both datasets. We attribute this feature to hardware/readout and tomography reconstruction imperfections, rather than entanglement, as the values appear consistent with systematic experimental errors.

\subsection{CHSH inequalities}

\begin{figure}
\centering
\includegraphics[width=0.5\textwidth]{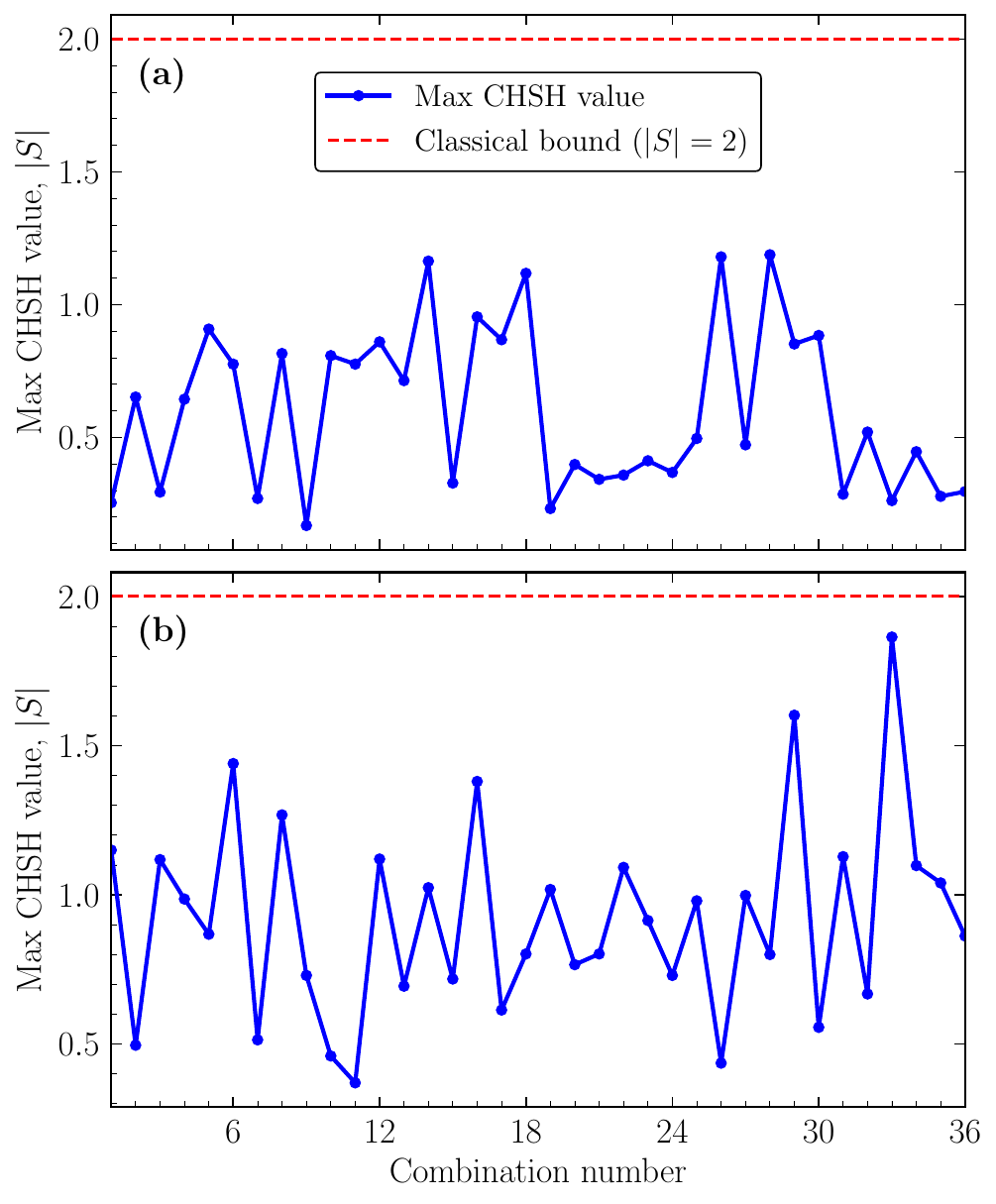}
\caption{Maximum value \(|S|\) of the CHSH inequality for the system of nuclear spin ($I = 1/2$) coupled to an NV center \textbf{(a)} and two coupled NV centers \textbf{(b)}. Each point represents the highest value of \(|S|\) reached when evaluating the correlations \(E(M,N)\) associated with the combination of considered measurements. The red--dashed horizontal line at $|S|=2$ indicates the classical bound.}
\label{fig:CHSH}
\end{figure}

Finally, CHSH inequalities are a standard tool used to test non--local correlations, since their violation rules out local hidden--variables models \cite{Clauser1969, Brunner2014, Werner1989}. The mathematical form of the CHSH inequality is $|S| \leq 2$
where
\be
S = E(A,B) - E(A,B') + E(A',B) + E(A', B').
\ee 
The terms $E(M,N)$, where \(M\in(A,A')\) and \(N \in (B, B')\), represent the expected value of the joint state for the measurement M in qubit 0 and N in qubit 1. The variables $A$, $A'$, $B$, $B'$ denote the measurement settings applied to the two systems in the experiment. In this case, these variables correspond to the Pauli measurements X, Y and Z for each qubit. The expected values may be expressed in terms of probabilities of measuerement outcomes $p(M=a,N=b)$
\bea
E(XX) &=& p(M=0,N=0) + p(M=1,N=1) \\
\nonumber &&- p(M=0,N=1)- p(M=1,N=0).
\eea
Since $A$\(\neq\)$A'$ and $B$\(\neq\)$B'$ and we performed meaurements on 3 the Pauli matrices (X, Y, Z), we evaluate $S$ over the 36 valid combinations of settings. To visualize the results, the graph shows the maximum CHSH value \(|S|\) at any given delay time obtained for each of the 36 combinations of measurements tested (XX, XY, XZ, etc.) \cite{Hensen2015, Aspect1982}.
Both for the case of the NV-nuclear spin system and the NV-NV system, as can be observed in Fig. \ref{fig:CHSH}, all the CHSH values remain below the threshold $|S| = 2$. In the first case \textbf{(a)}, this definitely rules out the presence of quantum entanglement. In the second case \textbf{(b)}, although entanglement is present, it is not sufficiently strong to grant violations of the CHSH inequalities. Nevertheless, several combinations, specifically combination 33, approach the bound, suggesting the presence of strong correlations that nearly saturate the inequality.

\section{Conclusion}
\label{sec:sec7}

We have validated and demonstrated the effectiveness of using quantum computers based on superconducting circuits as a versatile platform for simulating and modeling spin environments of NV-based quantum sensors.
Through the implementation of Hahn echo sequences and quantum state tomography (QST), we have been able to accurately characterize how different types of impurities affect sensor coherence and produce effects such as entanglement.
The analysis of decoherence revealed two distinct regimes. In the case of NV--nuclear spin interaction, we observed a moderate reduction of the coherence time (T2), consistent with a quasi-static model and the spectator decay induced dephasing (SDID) mechanism.
In contrast, in the NV--NV interaction, a drastic reduction in coherence is introduced as well as the appearance of oscillations in the dynamics of coherence. These enhanced results suggest the possible presence of entanglement, in addition to the aforementioned SDID effect.

An essential contribution of this work lies in the verification of the quantum nature of the interactions by reconstructing the density matrix. Purity analysis and the Peres--Horodecki criterion demonstrated that, while the NV-nuclear spin configuration evolves as a separable state, the NV-NV configuration generates entanglement, evidenced by the presence of negative eigenvalues in the partially transposed matrix.

It is important to note that, although the presence of entanglement was confirmed using the PPT criterion, no violation of the CHSH inequalities was observed in any of the scenarios. This finding suggests that, although the NV-NV system exhibits non--classical correlations, the generated state does not reach the non--locality threshold required to violate the classical limit under the current experimental conditions, possibly due to the rapid decoherence inherent in the hardware or the fidelity of the applied quantum gates, considering the latter as the main reason motivated by initial purity values below 1.

In summary, this work has demonstrated the use of quantum computers as an accessible tool to model quantum processes such as the dynamics of nano--scale sensors.

\section{Acknowledgments}

A.L.G and J.C. acknowledge support from grant CNS2023-144994 funded by MICIU/AEI/10.13039/201100011033 and by ``ERDF/EU''. J.C. additionally acknowledges support from European Union project C-QuENS (Grant No. 101135359).

\bibliography{NVQ3} 
\bigskip  

\newpage


\bigskip  

\appendix

\end{document}